# Transport properties of the pseudospin-3/2 Dirac--Weyl fermions in the double-barrier-modulated two-dimensional system

Rui Zhu[1]

School of Physics and Optoelectronics, South China University of Technology,
Guangzhou 510641, People's Republic of China

**Abstract**

The pseudospin-3/2 Dirac--Weyl system is the kind of system bearing the quasiparticle band structure of two cones with different apex angles and their reversed replica touching at the apex, whose properties can be described by the pseudospin-3/2 Dirac equation. In this work, we analytically solved the pseudospin-3/2 Dirac equation and investigated the electronic transport properties in the double-barrier modulated two-dimensional system. The probability current density operator is explicitly derived from the time-dependent pseudospin-3/2 Dirac equation, which paves way for investigation of the electronic transport properties of general pseudospin-$s$ Dirac--Weyl systems with $s$ an integer or half integer larger than 1. As a result of the double-cone band structure, the pseudospin-3/2 system has two incident channels for a single incident energy and incident angle pair. Similar to its counterparts of pseudospin-1/2 and pseudospin-1 Dirac--Weyl systems, the Klein tunneling and resonant tunneling effects in the transmission probability are numerically observed for incidence coming from both Dirac cones in the double-barrier-modulated pseudospin-3/2 system. In contrast to its pseudospin-1/2 and -1 counterparts, the Klein tunneling and resonant tunneling effects are differentiated into double-channel and single-channel incidences, corresponding to different regimes in the $E$-$k_y$ dispersion plane. Without a flat band, the super Klein tunneling effect of the pseudospin-1 Dirac--Weyl system does not occur in the pseudospin-3/2 system. Using the numerically obtained probability current density, the zero-temperature conductivity, shot noise, and Fano factor are calculated. As a combined result of double-channel incidence, Klein tunneling, and resonant tunneling, in comparison with its pseudospin-1/2 (graphene) and pseudospin-1 counterparts, the conductivity and shot noise in the pseudospin-3/2 double-barrier structure is enhanced. A Fano factor between 0.4 and 0.5 close to the Dirac point $E_F = V_0$ is observed.

---

[1] Corresponding author. Electronic address: rzhu@scut.edu.cn

# I. INTRODUCTION

In investigation of the transport properties, it is found that the band structure of the electronic states plays an important role including the parabolic, linear or other configurations of dispersions [1-6]. Along with the technical development in material fabrication and simulation, the lattice structure within the capacity of both experimental and theoretical researches becomes more and more complicated and sophisticated. It is known in theory that the number of the energy bands for a single momentum in the momentum space of the single-particle electronic states of a certain crystal is equal to that of the atomic orbitals within one unit cell under the Bloch's theorem, which is also the dimension of the $\text{Sublattice} \otimes \text{Orbital}$ Hilbert space irrespective of the spin degree of freedom. Almost chronologically in the context of quantum transport, physicists witness the "evolution" of the emerging research targets from the cubic lattice to the graphene [2,3,7-17]; the Lieb, Dice or $T_3$ Kagame, and $K_4$ lattices [18-27]; antiperovskites [28,30,63]; and etc. Thence, when the multiple parabolic electronic bands cross, close to the crossing point we have an effective $\mathbf{k} \cdot \mathbf{p}$ Hamiltonian of pseudospin-1/2, 1, 3/2, and etc., corresponding to the two-, three-, four-fold degenerate point respectively [31-61]. Quasiparticles in these Hamiltonians are called pseudospin-$s$ ($s = 1/2, 1, 3/2, \cdots$) Dirac--Weyl fermions or the Rarita-Schwinger-Weyl fermions [59,62] in the case of $s = 3/2$. Besides condensed matter systems, the pseudospin-$s$ fermionic or bosonic model is also relevant in various other physical scenarios when the quasiparticle band structure is analogous, such as the optical-lattice-confined ultracold atoms [43,45,46], photonic crystals [37-40], and etc. Previously, physical properties of the pseudospin-1/2 and -1 Dirac--Weyl system have been intensively investigated since its proposition. Recently, the higher pseudospin-$s$ systems with half-integer $s$ and $s \geq 3/2$ has attracted interest in the condensed-matter physical community due to the discovery of its material host and its remarkable topological nontriviality [6,28,59,63].

However, after the proposition of the higher pseudospin Dirac--Weyl fermions almost a decade ago, the main focus of the system is its nontrivial topology in the case of insulator, semimetal, metal, and superconductor, with and without consideration of the Coulomb interaction. A systematic investigation of the transport properties of the pseudospin-$s$ fermions with $s \geq 3/2$ attracts little interest among the physical community in spite of its significance both fundamentally, technically, and in application. One of the reasons for its lack of focus probably is that one might assume that the extension of the transport properties from the pseudospin-1/2 and 1 should be natural and straightforward. Nevertheless, after we undertook the task, we found that the quantum transport problem of the pseudospin-$s$ fermion with $s = 3/2$ is remarkably different from its $s = 1/2$ and 1 counterparts. Originating from the four-fold energy degenerate point, two difficulties come into attention.

One is that as a result of the doubly-degenerate conduction band, the incident, transmission, and reflection channel simultaneously bifurcates into two. In such a situation, the definition of the transmission probability and unitarity of the scattering matrix is unprecedented.

The other is that it is known that under modulation of the electrostatic potential barrier, the Klein tunneling effect occurs in the graphene and the pseudospin-1 Dirac--Weyl fermionic system when chirality matches between the electronic states outside of the barrier (conduction band) and the hole states inside the barrier (valence band). And in the case of double barriers, resonant tunneling effect occurs in the regime of the energy-momentum space where we have electronic states outside of the barrier (conduction band) and forbidden states in the barrier region (outside of both the conduction and valence bands like in the band gap of the conventional parabolically-dispersive semiconductor). In contrast to its pseudospin-1/2 and -1 counterparts, in the pseudospin-3/2 fermionic system, the quasiparticle energy dispersion has double cones with different slope angles for both the conduction and valence bands. Hence, the problem is in which regime of the energy-momentum space should the Klein tunneling and resonant tunneling effects occur and how should one understand the quantum mechanisms of these two effects.

In solution to the first problem, we transplanted the higher-spin Dirac theory in particle physics to the higher-pseudospin model in the condensed matter physics. We derived the probability current density operator from the time-dependent (pseudo)spin-3/2 Dirac equation and continuity equation of the particle number conservation. Using the probability current density operator, we obtained the well-defined transmission and reflection probabilities with simultaneous involvement of the transmission and reflection amplitudes in both scattering channels corresponding to the energy-degenerate double Dirac cones. And by numerical confirmation of the conservation of the probability current in different scattering regions, the problem of the unitarity of the scattering matrix is solved.

In solution to the second problem, we found four regimes in the $E$-$k_y$ dispersive space corresponding to the occurring condition of the double-channel Klein tunneling, single-channel Klein tunneling, double-channel resonant tunneling, single-channel resonant tunneling effects, respectively. We know that all the scattering channels belong to one spinor wavefunction obtained from the pseudospin-3/2 Dirac--Weyl Hamiltonian. Definition of the four different regimes is more important in the sense of technique than in the sense of concept. However, because it is technically meaningful, one cannot skip the channel discussions to solve the transport problem and understand the transport properties of the pseudospin-3/2 Dirac--Weyl fermions. Details of the four transport regimes and discussions about their collaborative contribution to the conductivity will be presented in the following other sections of the paper.

By extension of the pseudospin-3/2 Dirac equation to a general pseudospin-$s$ Dirac equation with arbitrary $s > 3/2$, our solution to the above-introduced two difficulties paves way to investigate and understand the quantum transport properties of the electronic states in a general pseudospin-$s$ Dirac--Weyl fermionic system with higher pseudospins. In this work, we calculated the transmission probability, conductivity, shot noise, and Fano factor of the pseudospin-3/2 Dirac--Weyl fermions in the two-dimensional (2D) system subjected to modulation of double electrostatic potential barriers. The results demonstrate transport properties characteristic of the pseudospin-3/2 Dirac--Weyl fermions. Although no direct phenomenon of the nontrivial topology like quantized conductivity and Fano factor is found, the results of Klein tunneling and resonant tunneling effects unique for pseudospin-3/2 fermions will lend insight to the experimental research and potential applications and indirectly show signatures of their nontrivial topology such as from the relation between the resonant energy and the quasibound discrete levels. The rest of this paper is organized as follows. Sec. 2 is attributed to the model and formalisms. In Sec. 3, numerical results are presented. Sec. 4 is devoted to a brief summary.

## II. MODEL AND FORMALISM

The 2D pseudospin-3/2 Dirac--Weyl band structure is given by two reversed pairs of different-slope Dirac cones touching at the apices of the cones as shown in Fig. 1. The low-energy effective Hamiltonian originating from the four-fold degenerate $\Gamma$ point of the antiperovskite $A_3BX$ crystal is governed by an effective Hamiltonian of pseudospin-3/2 Dirac--Weyl form assuming that we could obtain a 2D film including the $\Gamma X$ line in the momentum space from the three-dimensional antiperovskite crystal [28,59]

$$\hat{H} = -i\hbar v_g \hat{\mathbf{S}} \cdot \nabla + V(x), \tag{1}$$

Here, $v_g$ is the group velocity of the quasiparticles. $\hat{\mathbf{S}} = (\hat{S}_x, \hat{S}_y, \hat{S}_z)$ is the spin-3/2 operator in the $\hat{S}_z$-representation (the representation that $\hat{S}_z$ is diagonalized) with

$$\hat{S}_x = \begin{pmatrix} 0 & \frac{\sqrt{3}}{2} & 0 & 0 \\ \frac{\sqrt{3}}{2} & 0 & 1 & 0 \\ 0 & 1 & 0 & \frac{\sqrt{3}}{2} \\ 0 & 0 & \frac{\sqrt{3}}{2} & 0 \end{pmatrix}, \quad \hat{S}_y = \begin{pmatrix} 0 & -\frac{\sqrt{3}}{2}i & 0 & 0 \\ \frac{\sqrt{3}}{2}i & 0 & -i & 0 \\ 0 & i & 0 & -\frac{\sqrt{3}}{2}i \\ 0 & 0 & \frac{\sqrt{3}}{2}i & 0 \end{pmatrix},$$

$$\hat{S}_z = \frac{1}{2}\begin{pmatrix} 3 & 0 & 0 & 0 \\ 0 & 1 & 0 & 0 \\ 0 & 0 & -1 & 0 \\ 0 & 0 & 0 & -3 \end{pmatrix},$$

(2)

satisfying $[\hat{S}_k, \hat{S}_l] = i\varepsilon_{klm}\hat{S}_m$. In consideration of the 2D system, only $\hat{S}_x$ and $\hat{S}_y$ appear in the Hamiltonian (1). As in the conventional semiconductor double-barrier diodes, two symmetric electrostatic potential barriers are applied to the system with $V(x) = V_0$ in the barrier region and 0 outside of the barriers indicated in the device cartoon of Fig. 1. Hight of the barriers can be tuned by chemical doping or gate-voltage variation.

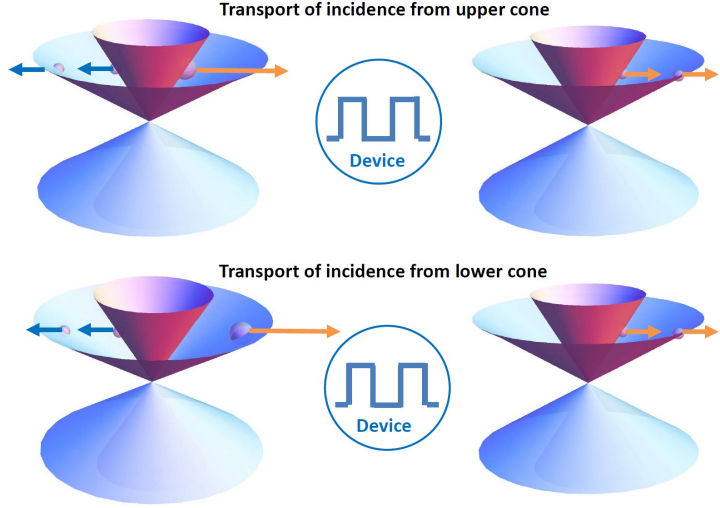

FIG. 1: Schematics of the band structure of the 2D pseudospin-3/2 fermionic system and tunneling channels in the transport process through a nano-device. Both the transmission and reflection channels bifurcate into two corresponding to the energy-degenerate double Dirac cones. Incidence can occur from either of the upper (larger-slope) or lower (smaller-slope) Dirac cones, constituting two independent transverse channels (when $k_y$ is fixed, $k_x$ and the incident angle are different for the two cones.) in consideration of the Landauer--Büttiker conductivity and shot noise. Physical picture of the scattering channels is relevant but not limited to tunneling through double barriers.

For $V=0$, the eigenvalues and the corresponding normalized (as shown below, due to the double-cone-degenerate band structure, the flux normalization does not apply to the present situation) eigen-spinors of the Hamiltonian (1) are

$$E_{1,4} = \pm \frac{3}{2}\hbar v_g \sqrt{k_x^2 + k_y^2}, \quad E_{2,3} = \pm \frac{1}{2}\hbar v_g \sqrt{k_x^2 + k_y^2}, \tag{3}$$

$$\psi_i = \varphi_i e^{ik_x x + ik_y y}, \quad i=1,2,3,4, \tag{4}$$

$$\begin{aligned}
\varphi_1 &= \left(e^{-i3\theta}, \ \sqrt{3}e^{-i2\theta}, \ \sqrt{3}e^{-i\theta}, \ 1\right)^T / \sqrt{8}, \\
\varphi_2 &= \left(-e^{-i3\theta}, \ -\frac{1}{\sqrt{3}}e^{-i2\theta}, \ \frac{1}{\sqrt{3}}e^{-i\theta}, \ 1\right)^T / \sqrt{\frac{8}{3}}, \\
\varphi_3 &= \left(e^{-i3\theta}, \ -\frac{1}{\sqrt{3}}e^{-i2\theta}, \ -\frac{1}{\sqrt{3}}e^{-i\theta}, \ 1\right)^T / \sqrt{\frac{8}{3}}, \\
\varphi_4 &= \left(-e^{-i3\theta}, \ \sqrt{3}e^{-i2\theta}, \ -\sqrt{3}e^{-i\theta}, \ 1\right)^T / \sqrt{8}.
\end{aligned} \tag{5}$$

Here, $\theta = \arctan(k_y/k_x)$ and 1 to 4 labels decreasingly ordered four eigenvalues. We consider an infinitely-wide sample and $k_y$ is conserved during tunneling as a result of the translational symmetry in the transverse direction. (Below we see that in the transport consideration, for a certain pair of $E$ and $k_y$, $k_x$ and $\theta$ are different for $i=1,4$ from $i=2,3$.) For $V=V_0$, each eigenvalue is directly added by a $V_0$ with the form of the eigen-spinors unchanged and $k_x$ changed, the latter of which can be obtained from Eq. (3) for a certain $k_y$.

As in a usual transport approach, we start from the spinor wavefunction in different scattering regions and the continuity relation at the interfaces. Here, we should note and be careful that there are two different and independent incident channels for a fixed pair of incident energy $E$ and transverse momentum $k_y$ and two transmission and two reflection amplitudes for each incident channel as a result of the double-cone degeneracy as illustrated in Fig. 1. The spinor wavefunction for incidence from the upper Dirac cone when $E < V_0$ is

$$\Psi_H = e^{ik_y y} \begin{cases} \varphi_{1\rightarrow} e^{ik_{1x}x} + r_{11}\varphi_{1\leftarrow} e^{-ik_{1x}x} + r_{21}\varphi_{2\leftarrow} e^{-ik_{2x}x}, & x<0, \\ A_{3\rightarrow}\varphi_{3\rightarrow} e^{i\kappa_{3x}x} + A_{3\leftarrow}\varphi_{3\leftarrow} e^{-i\kappa_{3x}x} \\ +A_{4\rightarrow}\varphi_{4\rightarrow} e^{i\kappa_{4x}x} + A_{4\leftarrow}\varphi_{4\leftarrow} e^{-i\kappa_{4x}x}, & 0<x<L, \\ B_{1\rightarrow}\varphi_{1\rightarrow} e^{ik_{1x}x} + B_{1\leftarrow}\varphi_{1\leftarrow} e^{-ik_{1x}x} \\ +B_{2\rightarrow}\varphi_{2\rightarrow} e^{ik_{2x}x} + B_{2\leftarrow}\varphi_{2\leftarrow} e^{-ik_{2x}x}, & L<x<L+W, \\ C_{3\rightarrow}\varphi_{3\rightarrow} e^{i\kappa_{3x}x} + C_{3\leftarrow}\varphi_{3\leftarrow} e^{-i\kappa_{3x}x} \\ +C_{4\rightarrow}\varphi_{4\rightarrow} e^{i\kappa_{4x}x} + C_{4\leftarrow}\varphi_{4\leftarrow} e^{-i\kappa_{4x}x}, & L+W<x<2L+W, \\ t_{11}\varphi_{1\rightarrow} e^{ik_{1x}x} + t_{21}\varphi_{2\rightarrow} e^{ik_{2x}x}, & x>2L+W. \end{cases} \qquad (6)$$

The spinor wavefunction for incidence from the lower Dirac cone when $E<V_0$ is

$$\Psi_L = e^{ik_y y} \begin{cases} \varphi_{2\rightarrow} e^{ik_{2x}x} + r_{12}\varphi_{1\leftarrow} e^{-ik_{1x}x} + r_{22}\varphi_{2\leftarrow} e^{-ik_{2x}x}, & x<0, \\ \cdots, & 0<x<2L+W, \\ t_{12}\varphi_{1\rightarrow} e^{ik_{1x}x} + t_{22}\varphi_{2\rightarrow} e^{ik_{2x}x}, & x>2L+W. \end{cases} \qquad (7)$$

$\Psi_H$ and $\Psi_L$ have the same form in the middle region $0<x<2L+W$ with different values of the coefficients. Numbers in the subscripts share the same definition as in the eigenvalues and eigen-spinors. We set the longitudinal size of the device $D=2L+W$ with $L=D/4$ and $W=D/2$. It is convenient to express length in the unit of $D$ and energies in units of $\hbar v_g/D$. Following this unit setting, in Eqs. (6) and (7), $k_{1x}=\sqrt{(2E/3)^2-k_y^2}$, $k_{2x}=\sqrt{(2E)^2-k_y^2}$, $\kappa_{3x}=\sqrt{4(E-V_0)^2-k_y^2}$, $\kappa_{4x}=\sqrt{4(E-V_0)^2/9-k_y^2}$. Outside of the potential barriers (in regions $x<0$, $L<x<L+W$, and $x>2L+W$), $\varphi_{i\rightarrow}=\varphi_i$ of Eq. (5) and $\varphi_{i\leftarrow}$ is obtained by replacing $k_{ix}$ by $-k_{ix}$ into $\varphi_{i\rightarrow}$ in the same region. Inside of the potential barriers (in regions $0<x<L$ and $L+W<x<2L+W$), $\varphi_{i\rightarrow}$ is obtained by replacing $k_{ix}$ by $\kappa_{ix}$ into the $\varphi_{i\rightarrow}$ outside of the potential barriers and $\varphi_{i\leftarrow}$ is obtained by replacing $\kappa_{ix}$ by $-\kappa_{ix}$ into the $\varphi_{i\rightarrow}$ in the same region.

The continuity relation at the interfaces can be obtained by integrating the Dirac–Weyl equation $\hat{H}\Psi=E\Psi$ over a small interval traversing the interface. The result is that each of the four components of the spinor wavefunction be continuous respectively. Two sets of the continuity relation are applied to $\Psi_H$ and $\Psi_L$ independently to obtain all the eight scattering amplitudes--- $r_{11}$, $r_{21}$, $t_{11}$, and $t_{21}$ for incidence from the upper cone in $\Psi_H$; $r_{12}$, $r_{22}$, $t_{12}$, and $t_{22}$ for incidence from the lower cone in $\Psi_L$. Here, $r_{ji}/t_{ji}$ ( $i,j=1,2$ ) is defined as the reflection/transmission amplitudes from the scattering channel $i$ to the scattering channel $j$. 1 and 2 correspond to the upper and lower cones respectively.

In the transport consideration based on the Schrödinger equation, one uses the continuity equation of the particle number conservation

$$\frac{\partial \rho}{\partial t} + \nabla \cdot \mathbf{j} = 0, \tag{8}$$

where $\rho$ is the probability density and $\mathbf{j}$ is the current density, together with the time-dependent Schrödinger equation to obtain the current density operator $\mathbf{j}$ and hence the transmission probability, conductivity, and other transport properties. In the case of the infinitely-wide model, only the component in the transport direction of the current operator matters, which is the $x$-component $j_x$ in our consideration. Here, analogously, we use the continuity equation (8) and the time-dependent (pseudo)spin-3/2 Dirac equation to obtain the current density operator. The time-dependent (pseudo)spin-3/2 Dirac equation is [64]

$$i\hbar \frac{\partial \Psi}{\partial t} = -i\hbar v_g \mathbf{S} \cdot \nabla \Psi, \tag{9}$$

where $\Psi$ is the spinor wavefunction and $\mathbf{S}$ is the spin-3/2 matrix defined in Eq. (2). Here we omit $V(x)$ in the Hamiltonian (1) because the current density is measured outside of the device, wherein $V(x) = 0$ in our approach. Substituting the time-dependent (pseudo)spin-3/2 Dirac equation (9) and $\rho = \Psi^\dagger \Psi$ into the continuity equation (8) and after some algebra we obtain the current density operator in the $x$-direction as

$$j_x = v_g \Psi^\dagger \hat{S}_x \Psi, \tag{10}$$

where $\hat{S}_x$ can be obtained in Eq. (2). Obviously, this formula is correct in dimension equality. Substituting the above obtained spinor wavefunctions $\Psi_H$ and $\Psi_L$ in Eqs. (6) and (7) into Eq. (10), one can explicitly obtain the analytic expressions of the incident, reflected, and transmitted current fluxes in the $x$-direction, respectively.

It is noted here that $\Psi_H$ and $\Psi_L$ constitute two independent incident channels and generate the current density fluxes independently. In consideration of the conductivity, shot noise, and other transport properties, we simply add the two channels analogous to what we manage the other kinds of independent channels. Conveniently, we label the transport quantities related to the two channels by a subscript $H$ or $L$, corresponding respectively to incidence from the upper-cone band (the one with larger slope) and to incidence from the lower-cone band (the one with smaller slope).

Substituting the incident term in the spinor wavefunctions Eq. (6)/(7) into the current density operator (10), we obtain the incident current density from the upper/lower cone

$$j_{xiH} = \frac{3}{2} v_g \cos\theta_H, \quad j_{xiL} = \frac{1}{2} v_g \cos\theta_L. \tag{11}$$

Here, $\theta_H$ and $\theta_L$ are different as the quasiparticle energy $E$ and the transverse momentum

$k_y$ are conserved during tunneling, which are defined as $\theta_H = \arctan(k_y/k_{1x})$ and $\theta_L = \arctan(k_y/k_{2x})$, respectively. Volume of the space is omitted in Eq. (11), which would be canceled out in calculation of the transmission probability and the conductivity. Substituting the reflection and transmission terms in the spinor wavefunctions Eqs. (6) and (7) into the current density operator (10), the reflected and transmitted current densities $j_{xrH/xrL}$ and $j_{xtH/xtL}$ can be obtained. To avoid tediousness, we do not provide the explicit expressions of the transmitted and reflected current densities $j_{xtH/xtL}$ and $j_{xrH/xrL}$ here. As in the conventional transport consideration, the strength of incident flux is set to unity to achieve a unitary scattering matrix, we can define the transmission and reflection probabilities as

$$T_{H/L} = \frac{\mathrm{Re}\, j_{xtH/xtL}}{\mathrm{Re}\, j_{xiH/xiL}}, \quad R_{H/L} = -\frac{\mathrm{Re}\, j_{xrH/xrL}}{\mathrm{Re}\, j_{xiH/xiL}}. \tag{12}$$

Here, several points are worth noting. One, in principle, the current density is a real quantity. But in the numerical treatment there is a vanishing imaginary part in it. The operation of Re is to secure accuracy. Two, because the derivation of the current density starts from the continuity equation, conservation of the current is naturally secured, i.e., we have $T + R = 1$, which is numerically confirmed. Three, two incident channels (incidence from the upper and lower cones as shown in Fig. 1) contribute independently to the transport process. The transmission and reflection probabilities as well as the unitarity of the scattering matrix all behave independently. Four, because the direction of the reflected current is opposite to the incident and transmitted currents, to have a positive reflection probability there is a minus sign in the right equation of Eq. (12).

With the transmission probability obtained, the conductivity $\sigma$ and shot noise $S$ can be calculated using the Landauer-Büttiker formalisms with [65]

$$\sigma = \frac{2e^2 D^2 V_0}{h^2 v_g}[\int_{-k_{1F}}^{k_{1F}} T_H(E_F, k_y) dk_y + \int_{-k_{2F}}^{k_{2F}} T_L(E_F, k_y) dk_y], \tag{13}$$

and

$$S = \frac{4e^3 D^2 V_0}{h^2 v_g}[\int_{-k_{1F}}^{k_{1F}} T_H(E_F, k_y)(1 - T_H(E_F, k_y)) dk_y \\ + \int_{-k_{2F}}^{k_{2F}} T_L(E_F, k_y)(1 - T_L(E_F, k_y)) dk_y]. \tag{14}$$

The Fano factor comparing the real shot noise and the Poisson noise is given by

$$F = S/(2e\sigma). \tag{15}$$

In Eqs. (13) and (14), because of the double-degenerate-cone band structure, for a certain Fermi energy $E_F$, the Fermi wave vector $k_F$ are different for incidence from the two different-slope

Dirac cones, which are $k_{1F} = 2E_F/3$ and $k_{2F} = 2E_F$, corresponding to the upper-cone and lower-cone incident channels, respectively.

Before we go on to present the numerical results of the above equations, the last part of this section is dedicated to illustrating by band structure the occurrence of the Klein tunneling and resonant tunneling effects, two characteristic phenomena previously observed in the pseudospin-1/2 and pseudospin-1 Dirac--Weyl systems, in the double-barrier-modulated 2D pseudospin-3/2 Dirac--Weyl system.

In the double-barrier-modulated graphene (pseudospin-1/2) and pseudospin-1 Dirac--Weyl system, the Klein tunneling effect occurs in the $E$-$k_y$ regime corresponding to the electronic states outside the barriers as well as hole states inside the barriers when perfect transmission probability appears under condition of chirality matching between the electronic and hole states [2,4,58]. Here in the case of the pseudospin-3/2 Dirac--Weyl fermions, the situation is different. For the illustration, Energy dispersions outside and inside of the electrostatic potential barriers are shown in Fig. 2. We know that propagating modes correspond to electronic states in the conduction bands as well as hole states in the valence bands. Here, we have double-cone propagating modes and single-cone propagating modes, the former of which exist above the upper cone (conduction band) and below the reversed upper cone (valence band); the latter of which exist above the lower cone and below the upper cone (conduction band) and above the reversed upper cone and below the reversed lower cone (valence band). As have illustrated, incidence from the two different-slope cones constitute two different conductive channels. Conveniently, we differentiate the Klein tunneling effect involving both-cone incidence from the Klein tunneling effect involving single-cone (only the lower cone) incidence. The occurring regions of the former and the latter are indicated as A and B, respectively in Fig. 2.

In the double-barrier-modulated graphene and pseudospin-1 Dirac--Weyl system, the resonant tunneling effect occurs when the propagating quasiparticle states coincide with the discrete quasibound states in the quantum well intersected between the two barriers whereas in the barriers the quasiparticle states are forbidden to propagate like in the energy gap of conventional semiconductors\cite{PereiraAPL2007, PereiraPRB2006, ZhuPhysicaB2024}. Here, in the case of the pseudospin-3/2 Dirac--Weyl fermions, the forbidden regime of quasiparticle propagation in the barrier region exists outside of all the four Dirac cones---two cones of the conduction band and two cones of the valence band, which constitutes regime B and D in Fig. 2, because when one channel is forbidden and the other is propagating, tunneling occurs and the state in the quantum well cannot be a bound state. Analogous to the Klein tunneling effect, the resonant tunneling process in the pseudospin-3/2 system can be differentiated into types B and D, corresponding to the double-cone incidence and single-cone incidence, as cartooned in Fig. 2.

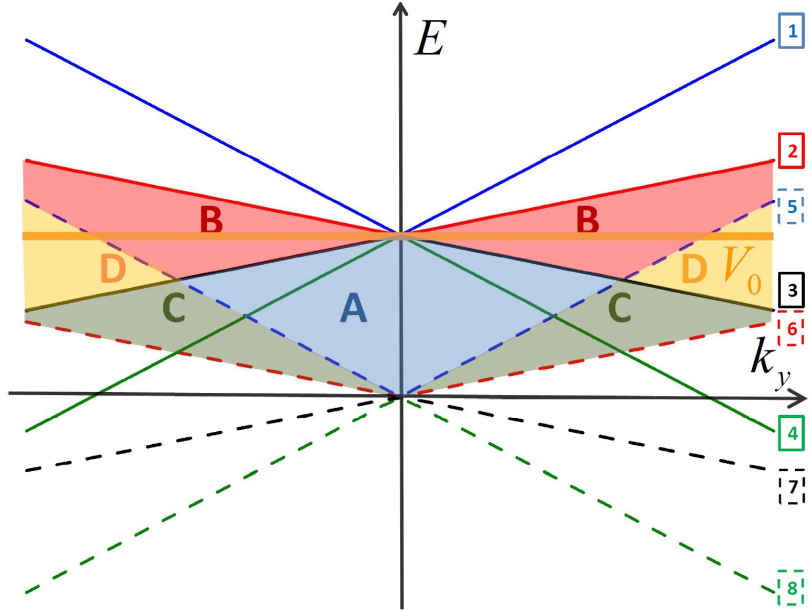

FIG. 2: Energy dispersions and occurring regions of the four types of the tunneling processes in the double-barrier-modulated 2D pseudospin-3/2 Dirac–Weyl system. Energy dispersions of the four bands—two conduction cone bands (blue and red in color) and two valence cone bands (black and olive in color)—in the barrier regions are indicated by solid-framed numbers 1 to 4. Energy dispersions of the four bands outside of the barriers are indicated by dashed-framed numbers 5 to 8. Correspondent bands inside of and outside of the barriers bear identical colors. The occurring regions of the four types of the tunneling processes—double-channel Klein tunneling, double-channel resonant tunneling, single-channel Klein tunneling, and single-channel resonant tunneling, are indicated by A, B, C, and D, respectively. Detailed discussions of the four types of the tunneling processes are in the main text.

Analogous to the resonant tunneling effect in other double-barrier-modulated systems, the quasibound states confined between the two barriers can be approximately obtained from the quasibound states confined in the finite-depth single quantum well with the well depth and width equal to the barrier height and the middle-region width in between the two barriers. The quasibound energy of the single finite-depth quantum well based on the pseudospin-3/2 model can be obtained from the solvability of the continuity relation at the well boundaries of the spinor wavefunction

$$\Psi_{QW} = e^{ik_y y} \begin{cases} D_{3\leftarrow}\varphi_{3\leftarrow}e^{-i\kappa_{3x}x} + D_{4\leftarrow}\varphi_{4\leftarrow}e^{-i\kappa_{4x}x}, & x<0, \\ F_{1\rightarrow}\varphi_{1\rightarrow}e^{ik_{1x}x} + F_{1\leftarrow}\varphi_{1\leftarrow}e^{-ik_{1x}x} \\ +F_{2\rightarrow}\varphi_{2\rightarrow}e^{ik_{2x}x} + F_{2\leftarrow}\varphi_{2\leftarrow}e^{-ik_{2x}x}, & 0<x<W, \\ G_{3\rightarrow}\varphi_{3\rightarrow}e^{i\kappa_{3x}x} + G_{4\rightarrow}\varphi_{4\rightarrow}e^{i\kappa_{4x}x}, & x>W. \end{cases} \quad (16)$$

As discussed above, the quasibound states exist in the forbidden region outside of all the four Dirac-cone bands in the regions out of the single quantum well. In this parameter regime both $\kappa_{3x}$ and $\kappa_{4x}$ are imaginary guaranteeing exponentially decaying wavefunctions of both the two channels to the leftward and rightward infinity outside the well region.

By categorization of the transport processes as types A, B, C, and D, difference between the pseudospin-3/2 Dirac--Weyl fermion and its pseudospin-1/2 and pseudospin-1 counterparts is clarified. It lends technical convenience to discussing numerical results of the transmission probability and calculating the conductivity and shot noise. It also paves way for analyzing the Klein tunneling and resonant tunneling effects in general higher pseudospin-$s$ ($s = 2, 5/2, \cdots$) Dirac--Weyl systems.

## III. RESULTS AND DISCUSSIONS

Numerical results of the transmission probability in the double-barrier-modulated 2D pseudospin-3/2 Dirac--Weyl system are shown in Fig. 3. Parameter regimes corresponding to the occurrence of the Klein tunneling and resonant tunneling effects are cartooned by colored stripes and indicated by letters A, B, C, and D. For comparison between the resonant energy and the quasibound states, the quasibound levels obtained from Eq. (16) are indicated by red dots in each subfigure. Using the transmission probability, the conductivity, shot noise, and Fano factor as a function of the Fermi energy are calculated from Eqs. (13) to (15), the results of which are shown in Fig. 4. In Figs. 3 and 4, we set $V_0 = 20\hbar v_g / D$. Other parameters are provided in the figure captions.

Transmission probabilities as a function of $k_y$ for $E = 0.5V_0$ and $E = 0.9V_0$ are shown in Fig. 3 (a) and (b), respectively. Transmission probabilities as a function of $E$ for $k_y = 2/D$ and $k_y = 10/D$ are shown in Fig. 3 (c) and (d), respectively. In the figure, we see that $T_L$ is visible in the whole coordinate scale, but $T_H$ is visible only in part of the coordinate scale. This is because that as discussed in the previous section, the quasiparticle state of incidence from the lower cone is propagating in the regime $k_y < 2E$ corresponding to $T_L$, while the quasiparticle state of incidence from the upper cone is propagating in the regime $k_y < 2E/3$ corresponding to $T_H$. It can be seen from the panel (a) and (b) that the Klein tunneling effect at normal incidence ($k_y = 0$) occurs for both the upper-cone incidence and lower-cone incidence channels corresponding to the transport type A. It can be seen from the panel (c) that the Klein tunneling effect at oblique incidence (e.g., $k_y = 0.2V_0$ as shown in the figure) occurs for both the upper-cone incidence and lower-cone incidence channels also corresponding to the transport type A. As shown in the panel (a) with $E = V_0/2$, no super Klein tunneling effect occurs for either of the two channels probably due to the nonexistence of the flat band, which is a shared property with the pseudospin-1/2 system and different from the pseudospin-1 system, the latter of which hosts the super Klein tunneling effect. In panel (a), (c), and (d), sharp perfect transmission peaks appear corresponding to region C of Fig. 2 demonstrating single-channel Klein tunneling effect at oblique incidence when $k_y \neq 0$.

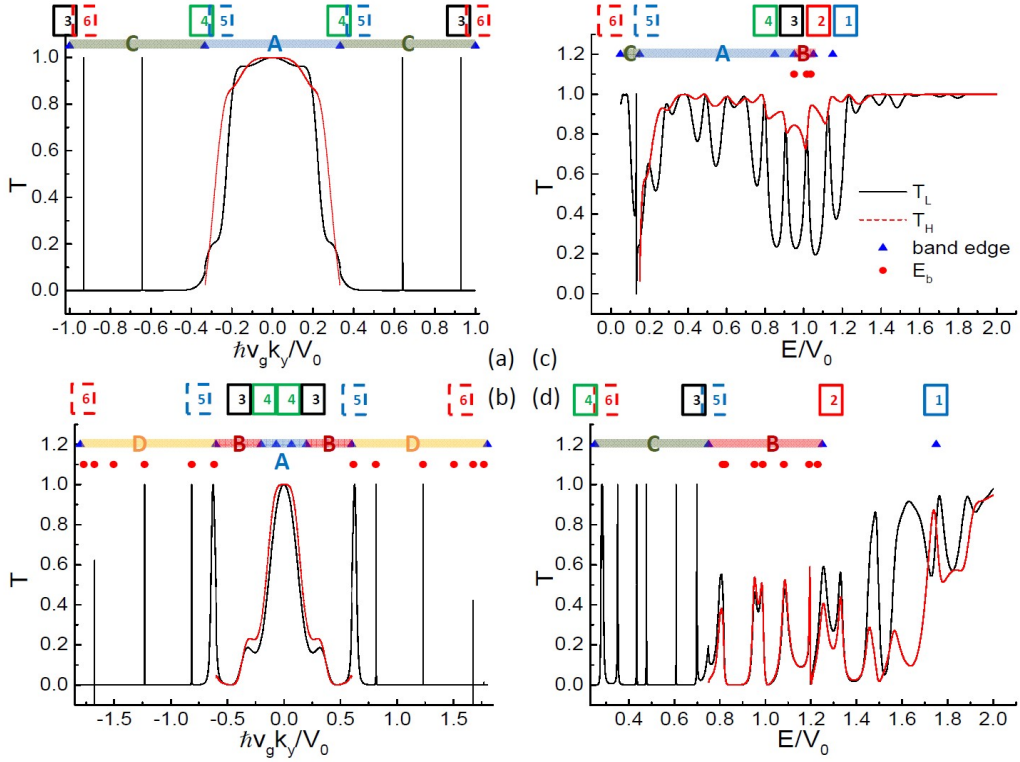

FIG. 3: Transport probability of the quasiparticles in the double-barrier-modulated 2D pseudospin-3/2 Dirac--Weyl system. (a)(b) Transport probability as a function of the transverse momentum $k_y$ for different incident energies $E$. $E = 0.5V_0$ in (a) and $E = 0.9V_0$ in (b). (c)(d) Transmission probability as a function of the incident energy $E$ for different transverse momentum $k_y$. $k_y = 2/D$ in (c) and $k_y = 10/D$ in (d). The band boundaries are indicated by blue triangles and framed numbers. The framed number refers to the particular band edge as applied in Fig. 2. Interlocked frames indicate overlap of band edges. The colored stripes connecting the band edges indicate the occurring regimes of different types of the tunneling processes as cartooned in Fig. 2 and discussed in the main text. Resonant tunneling occurs when the incident energy of the quasiparticle coincides with the discrete quasibound energy level $E_b$. In the figure, $E_b$ is indicated by red disks to show the energy coincidence of resonant tunneling effect. In obtaining the numerical results of panel (b), we found that deep inside the single-channel-resonant-tunneling region D ($k_y \gg 0$), the width of the resonance peak in energy is in the order of $10^{-7}V_0$ touching the limit of the numerical capacity. The unstable minute minus $T_L$ in the two outward resonant peaks of panel (b) is due to this reason.

Double-channel and single-channel resonant tunneling effect corresponding to type B and D respectively can be seen in Fig. 3 (b), (c), and (d). The quasibound levels of the single quantum well with identical depth and width with the well formed between the two barriers are marked by red dots. By comparison between the resonant-peak energy and the quasibound levels, their coincidence is obvious. In panel (d), we could see that in the type B regime, both $T_H$ and $T_L$ demonstrate prominent resonant peaks when the incident energy $E$ coincides with the quasibound level $E_b$. In panel (b), we could see that in the type D regime, the $T_L$ demonstrates prominent resonant peaks when the incident energy $E$ coincides with the quasibound level $E_b$. In panel (b), the resonance deep into the propagation-forbidden region (outside of all the four cone bands and $k_y \gg 0$) is weak in height or missing. It is because that deep into the forbidden region the quasibound state is so localized and difficult to connect with the incident propagating electronic wave, giving rise to a weak resonance with the peak width extremely small and beyond our numerical capacity (we found the peak width smaller than $10^{-7}V_0$ and the numerical results unstable in the next to last peak counting rightward). The parameter setting of panel (c) determines a narrow type B region. Only three quasibound levels exist within the corresponding quantum well, with two near the band edge. In this panel the energy correspondence between the resonant energy and the quasibound level is not strict. This is because that the situation of the double barrier structure is different from the single quantum well, in the latter of which we calculated the quasibound level $E_b$. As a result, the quasibound level in the double-barrier structure is slightly different, giving rise to the slight mismatch between the resonance energy and $E_b$.

In Fig. 4, the conductivity $\sigma$, shot noise $S$, and Fano factor $F$ as a function of the Fermi energy $E_F$ are shown. As discussed above, the quasiparticle energy dispersion of the pseudospin-3/2 Dirac--Weyl system has two different-slope cone-shape bands and their reverse replica sharing the same apex and symmetric axis. The remarkable transport feature resulting from such band structure is the existence of two independent incident channels corresponding to incidence from the two different-slope Dirac cones with the same quasiparticle energy $E$ and the transverse wave vector $k_y$. As a combined result of double-channel incidence, Klein tunneling, and resonant tunneling, in comparison with its pseudospin-1/2 (graphene) and pseudospin-1 counterparts, the conductivity and shot noise in the pseudospin-3/2 double-barrier structure are enhanced and demonstrate more local minimums and maximums as a function of the Fermi energy and no quantized conductivity minimum at the Dirac point of $E_F = V_0$ is observed. Due to the accumulative contribution from the transmission probability of all the transverse modes, the relation between the conductivity, shot noise, and Fano factor with the transmission probability of a certain $k_y$ is not obvious, as comparison between Figs. 3 and 4 suggests. However, as a result of the double-channel incidence, one can predict that it is possible that the Fano factor of the shot noise in the pseudospin-3/2 system present different properties from its pseudospin-1/2 and pseudospin-1 counterparts. From simple arithmetic analysis and previous knowledge of the shot noise in other systems [65], one can see that the the Fano factor of the shot noise contributed from two $T$, e.g., $T_1$ and $T_2$, varies in a large scale depending on the relative strength of $T_1$ and $T_2$. The Fano factor of the shot noise resulting from the combined effect of open channels with $T \approx 1$ and $F \approx 0$ (ballistic shot noise), closed channels with $T \approx 0$ and $F \approx 1$ (Poissonian shot noise), Pauli exclusion with $F$ approaching 1/2, and those channels with $0 < T < 1$, produces a Fano Factor of $F$ equal to or close to 1/3 in the single-barrier-modulated and double-barrier-modulated graphene [12,17], and a Fano facto of $F$ about 1/4 in the single-barrier-modulated and double-barrier-modulated pseudospin-1 Dirac--Weyl system [31,58] at or close to the Dirac point of $E_F = V_0$. Numerical results of the Fano factor of the shot noise of the double-barrier-modulated pseudospin-3/2 system demonstrates a value between 0.4 and 0.5 close to the Dirac point of $E_F = V_0$ as shown in Fig. 4. This remarkably different property from the graphene and pseudospin-1 system is well explained by the double-channel-incidence transport mechanism of the pseudospin-3/2 system. Its experimental observation will lend confirmation of the existence of the 2D pseudospin-3/2 Dirac--Weyl fermions in relevant materials and configurations, and suggest their potential exploitations.

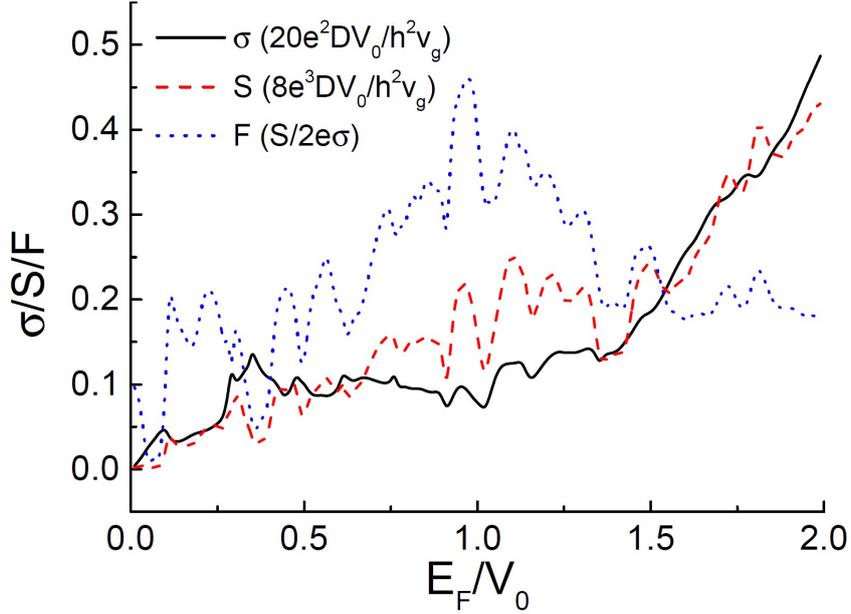

FIG. 4: Conductivity $\sigma$, shot noise $S$, and Fano factor $F$ as a function of the Fermi energy $E_F$ obtained from Eqs. (13), (14), and (15).

## IV. CONCLUSIONS

We summarize the main results of the present approach: One, from the continuity equation of quasiparticle number conservation and the time-dependent (pseudo)spin-3/2 Dirac equation, the analytical expression of the current density operator characterizing the transport properties of the pseudospin-3/2 Dirac--Weyl fermions is derived, which paves way for investigation of transport properties in the general higher pseudospin-$s$ ($s = 3/2, 2, \cdots$) Dirac--Weyl system. Two, the occurrence of the Klein tunneling and resonant tunneling effects in different regimes on the complex energy dispersion space of the double-barrier-modulated pseudospin-3/2 Dirac--Weyl system is clarified. Three, the Fano factor of the shot noise demonstrates a value between 0.4 and 0.5 close to the Dirac point of $E_F = V_0$, which differs remarkably from the graphene and pseudospin-1 Dirac--Weyl system subjected to the same electrostatic potential configuration. This result can be well explained by the unique band structure of the pseudospin-3/2 Dirac--Weyl system.

## V. ACKNOWLEDGEMENTS

This project was supported by the National Natural Science Foundation of China (No. 11004063) and the Fundamental Research Funds for the Central Universities, SCUT (No. 2017ZD099).

**Data Availability Statement** This manuscript has no associated data or the data will not be deposited. [Authors comment: All data included in this manuscript are available upon request by contacting with the corresponding author.]

1 S. Datta, *Electronic Transport in Mesoscopic Systems* (Cambridge University Press, Cambridge, 1995).

2 M. I. Katsnelson, K. S. Novoselov, and A. K. Geim, Nat. Phys. **2**, 620 (2006).

3 M.-H. Liu, J. Bundesmann, and K. Richter, Phys. Rev. B **85**, 085406 (2012).

4 H.-Y. Xu and Y.-C. Lai, Phys. Rev. B **94**, 165405 (2016).

5 J. D. Malcolm and E.J. Nicol, Phys. Rev. B **90**, 035405 (2014).

6 W. Yang, T. Xiang, and C. Wu, Phys. Rev. B **96**, 144514 (2017).

7 K. S. Novoselov, A. K. Geim, S. V. Morozov, D. Jiang, M. I. Katsnelson, I. V. Grigorieva, S. V. Dubonos, and A. A. Firsov, Nature (London) **438**, 197 (2005).

8 A. H. Castro Neto, F. Guinea, N. M. R. Peres, K. S. Novoselov, and A. K. Geim, Rev. Mod. Phys. **81**, 109 (2009).

9 K.-L. Chiu and Y. Xu, Phys. Rep. **669**, 1 (2017).

10 J. Milton Pereira, Jr., P. Vasilopoulos, and F. M. Peeters, Appl. Phys. Lett. **90**, 132122 (2007).

11 J. Milton Pereira, Jr., V. Mlinar, F. M. Peeters, and P. Vasilopoulos, Phys. Rev. B **74**, 045424 (2006).

12 J. Tworzyd{\l}o, B. Trauzettel, M. Titov, A. Rycerz, and C. W. J. Beenakker, Phys. Rev. Lett. **96**, 246802 (2006).

13 P. San-Jose, E. Prada, S. Kohler, and H. Schomerus, Phys. Rev. B **84**, 155408 (2011).

14 E. Prada, P. San-Jose, and H. Schomerus, Phys. Rev. B **80**, 245414 (2009).

15 R. Zhu and H. Chen, Appl. Phys. Lett. **95**, 122111 (2009).

16 R. Zhu, J.-H. Dai, and Y. Guo, J. Appl. Phys. **117**, 164306 (2015).

17 R. Zhu and Y. Guo, Appl. Phys. Lett. **91**, 252113 (2007).

18 M.R. Slot, T.S. Gardenier, P.H. Jacobse, G.C.P. van Miert, S.N. Kempkes, S.J.M. Zevenhuizen, C.M. Smith, D. Vanmaekelbergh, and I. Swart, Nat. Phys. **13**, 672 (2017).

19 L. Wang and D.-X. Yao, Phys. Rev. B **98**, 161403(R) (2018).

20 W. Beugeling, J. C. Everts, and C. M. Smith, Phys. Rev. B **86**, 195129 (2012).

21 F. Wang and Y. Ran, Phys. Rev. B **84**, 241103(R) (2011).

22 W. Häusler, Phys. Rev. B **91**, 041102(R) (2015).

23 J. D. Malcolm and E. J. Nicol, Phys. Rev. B **93**, 165433 (2016).

24 M. Tsuchiizu, Phys. Rev. B **94**, 195426 (2016).

25 D. F. Urban, D. Bercioux, M. Wimmer, and W. Häusler, Phys. Rev. B **84**, 115136 (2011).

26 M. Vigh, L. Oroszlány, S. Vajna, P. San-Jose, G. Dávid, J. Cserti, and B. Dóra, Phys. Rev. B **88**, 161413(R) (2013).

27 R. Zhu, Phys. Lett. A **383**, 684 (2019).

28 T. H. Hsieh, J. Liu, and L. Fu, Phys. Rev. B **90**, 081112(R) (2014).

29 T. Kawakami, T. Okamura, S. Kobayashi, and M. Sato, Phys. Rev. X **8**, 041026 (2018).

30 H. Isobe and L. Fu, Phys. Rev B **93**, 241113(R) (2016).

31 R. Zhu and P. M. Hui, Phys. Lett. A **381**, 1971 (2017).

32 X.-Y. Mai, Y.-Q. Zhu, Z. Li, D.-W. Zhang, and S.-L. Zhu, Phys. Rev. A **98**, 053619 (2018).

33 T. Louvet, P. Delplace, A. A. Fedorenko, and D. Carpentier, Phys. Rev. B **92**, 155116 (2015).

34 M. Orlita, D. Basko, M. Zholudev, F. Teppe, W. Knap, V. Gavrilenko, N. Mikhailov, S. Dvoretskii, P. Neugebauer, C. Faugeras, A.-L. Barra, and M. Potemski, Nat. Phys. **10**, 233 (2014).

35 W. C. Wareham and E. J. Nicol, Phys. Rev. B **108**, 085424 (2023).

36 A. Fang, Z. Q. Zhang, S. G. Louie, and C. T. Chan, Phys. Rev. B **93**, 035422 (2016).

37 D. Guzmán-Silva, C. Mejía-Cortés, M. A. Bandres, M. C. Rechtsman, S. Weimann, S. Nolte, M. Segev, A. Szameit, and R. A. Vicencio, New J. Phys. **16**, 063061 (2014).

38 S. Mukherjee, A. Spracklen, D. Choudhury, N. Goldman, P. Öhberg, E. Andersson, and R. R. Thomson, Phys. Rev. Lett. **114**, 245504 (2015).

39 R. A. Vicencio, C. Cantillano, L. Morales-Inostroza, B. Real, C. Mejía-Cortés, S. Weimann, A. Szameit, and M. I. Molina, Phys. Rev. Lett. **114**, 245503 (2015).

40 F. Diebel, D. Leykam, S. Kroesen, C. Denz, and A. S. Desyatnikov, Phys. Rev. Lett. **116**, 183902 (2016).

41 B. Dóra, I. F. Herbut, and R. Moessner, Phys. Rev. B **90**, 045310 (2014).

42 Z. Lin and Z. Liu, J. Chem. Phys. **143**, 214109 (2015).

43 J. Wang, S. Deng, Z. Liu, and Z. Liu, National Sci. Rev. **2**, 22 (2015).

44 A. D. Kovács, G. Dávid, B. Dóra, and J. Cserti, Phys. Rev. B **95**, 035414 (2017).

45 H. Xu and Y.-C. Lai, Phys. Rev. A **95**, 012119 (2017).

46 O. Dutta, A. Przysi\c{e}\.{z}na, and M. Lewenstein, Phys. Rev. A **89**, 043602 (2014).

47 E. Illes and E. J. Nicol, Phys. Rev. B **94**, 125435 (2016).

48 J. D. Malcolm and E. J. Nicol, Phys. Rev. B **94**, 224305 (2016).

49 E. Illes, J. P. Carbotte, and E. J. Nicol, Phys. Rev. B **92**, 245410 (2015).

50 J. L. Mañes, Phys. Rev. B **85**, 155118 (2012).

51 J. D. Malcolm and E. J. Nicol, Phys. Rev. B **92**, 035118 (2015).

52 W. Li, M. Guo, G. Zhang, and Y.-W. Zhang, Phys. Rev. B **89**, 205402 (2014).

53 G. Giovannetti, M. Capone, J. van den Brink, and C. Ortix, Phys. Rev. B **91**, 121417 (2015).

54 R. Shen, L. B. Shao, B. Wang, and D. Y. Xing, Phys. Rev. B **81**, 041410 (2010).

55 B. Dóra, J. Kailasvuori, and R. Moessner, Phys. Rev. B **84**, 195422 (2011).

56 S. M. Cunha, D. R. da Costa, J. Milton Pereira, Jr., R. N. Costa Filho, B. Van Duppen, and F. M.

Peeters, Phys. Rev. B **105**, 165402 (2022).

57 N. Weekes, A. Iurov, L. Zhemchuzhna, G. Gumbs, and D. Huang, Phys. Rev. B **103**, 165429 (2021).

58 R. Zhu, Physica B **687**, 416091 (2024).

59 I. Boettcher, Phys. Rev. B **102**, 155104 (2020).

60 M. Ezawa, Phys. Rev. B **94**, 195205 (2016).

61 I. Mandal, Phys. Lett. A **384**, 126666 (2020).

62 L. Liang and Y. Yu, Phys. Rev. B **93**, 045113 (2016).

63 T. Kawakami, T. Okamura, S. Kobayashi, and M. Sato, Phys. Rev. X **8**, 041026 (2018).

64 I. V. Ivanova, V. M. Shabaev, D. A. Telnov, and A. Saenz, Phys. Rev. A **98**, 063402 (2018).

65 Y. M. Blanter and M. Büttiker, Phys. Rep. **336**, 1 (2000).